\documentclass{llncs}
\usepackage{cite}
\usepackage{wrapfig}
\usepackage{epsfig}
\usepackage{amsmath}
\usepackage{color}
\usepackage{verbatim}
\usepackage{hyperref}
\usepackage{times}
\usepackage{ulem}    
\usepackage{enumerate}

\usepackage[linesnumbered,ruled,procnumbered,noend]{algorithm2e}
\usepackage{stmaryrd}
\usepackage{thmtools}
\usepackage{thm-restate}

\usepackage[version=0.96]{pgf}
\usepackage{tikz}
\usetikzlibrary{arrows,shapes,decorations.pathmorphing,automata,backgrounds,petri}

\usepackage{amssymb}
\usepackage{multirow}

\DeclareSymbolFont{largesymbolsA}{U}{txexa}{m}{n}
\SetSymbolFont{largesymbolsA}{bold}{U}{txexa}{bx}{n}
\DeclareFontSubstitution{U}{txexa}{m}{n}
\DeclareMathSymbol{\bigsqcupplus}{\mathop}{largesymbolsA}{"02}

\newcommand{\forget}[1]{}

\title{A Forward Simulation-Based Hierarchy of Linearizable Concurrent Objects} 
\author{Chao Wang\inst{1} \and Ruijia Li\inst{1} \and Yang Zhou\inst{1} \and Peng Wu\inst{2,3} \and Yi Lv\inst{2,3} \and Jianwei Liao\inst{1}, Jim~Woodcock\inst{1,4,5,6} \and Zhiming Liu\inst{1}} 
\institute{
  School of Computer and Information Science, Southwest University, Chongqing, China 
  \and 
   Key Laboratory of System Software, Institute of Software, CAS, Beijing, China
  \and
  University of Chinese Academy of Sciences, Beijing, China 
  \and
  State Key Laboratory of Intelligent Vehicle Technology, Chongqing, China
  \and
  Aarhus University, Aarhus, Denmark
  \and 
  Department of Computer Science, University of York, York, UK 
}

\begin{document}
\maketitle

\begin{abstract} 
In this paper, we systematically investigate the connection between linearizable objects and forward simulation.
We prove that the sets of linearizable objects satisfying wait-freedom (resp., lock-freedom or obstruction-freedom) form a bounded join-semilattice under the forward simulation relation, and that the sets of linearizable objects without liveness constraints form a bounded lattice under the same relation.
Thus, forward simulation is not only a proof technique for linearizability but also induces an algebraic hierarchy of linearizable objects.
As part of our lattice result, we propose an equivalent characterization of linearizability by reducing checking linearizability w.r.t. sequential specification $Spec$ into checking forward simulation w.r.t. an object $\mathcal{U}_{Spec}$. 

\medskip

We prove that the Herlihy-Wing queue is simulated by $\mathcal{U}_{Queue}$ with $Queue$ the sequential specification of the queue.
Thus, our object $\mathcal{U}_{Spec}$ can be used in the verification of linearizability.
To demonstrate the forward simulation relation between concrete linearizable objects, we prove that the time-stamped queue simulates the Herlihy-Wing queue, while the Herlihy-Wing queue cannot simulate the time-stamped queue.
All these three proofs have been machine-verified by Isabelle/HOL.

\end{abstract}

\section{Introduction } 
\label{sec:introduction}

High-performance concurrent data structures have been widely used in concurrent programs to leverage multi-core architectures. 
Linearizability \cite{DBLP:journals/toplas/HerlihyW90,DBLP:books/daglib/0020056,DBLP:conf/ppopp/VafeiadisHHS06,DBLP:journals/jacm/ValeSC24,DBLP:conf/esop/FilipovicORY09} is the \textit{de facto} correctness condition for concurrent objects. 
Intuitively, linearizability requires that each operation takes effect atomically at some time point between its invocation and response.

In this paper, we consider forward simulation, which is typically used in the verification of linearizability \cite{DBLP:conf/cav/BouajjaniEEM17,DBLP:conf/cav/AmitRRSY07,DBLP:conf/vmcai/Vafeiadis09,DBLP:journals/sttt/AbdullaHHJR17,DBLP:journals/pacmpl/JayantiJYH24}. 
In \cite{DBLP:phd/ethos/Vafeiadis08,DBLP:journals/toplas/LiangFF14,DBLP:conf/pldi/LiangF13}, forward simulations towards abstract objects are used together with separation logic to verify linearizability. 
Recently, Jayanti $\textit{et al.}$ proposed in \cite{DBLP:journals/pacmpl/JayantiJYH24} a forward-reasoning-based verification approach for linearizability, which is complete for all linearizable objects. 
As far as we know, these studies primarily focus on proving that specific abstract objects can simulate target concrete objects to demonstrate the latter's linearizability. 
We raise a general problem: is there any underlying structure to the forward-simulation relations among all linearizable objects, rather than just  simulations towards specific objects? 
Our answer is yes because we find that these forward-simulation relations form a bounded join-semilattice or a bounded lattice, depending on whether liveness constraints are taken into account.

A liveness property describes a condition under which method calls are guaranteed to complete in an execution. 
Wait-freedom \cite{DBLP:journals/toplas/Herlihy91,DBLP:conf/concur/LiangHFS13}, lock-freedom \cite{DBLP:books/daglib/0020056,DBLP:conf/concur/LiangHFS13}, and obstruction-freedom \cite{DBLP:books/daglib/0020056,DBLP:conf/concur/LiangHFS13} are typical liveness properties. 
Intuitively, wait-freedom requires that each method call return in a finite number of steps. 
Given a sequential specification $Spec$, a $WF$-$\preceq_{(c,r)}$-class is a set of linearizable and wait-free objects that simulate each other. 
We generate a partial order $\preceq_{(c,r)}$ between $WF$-$\preceq_{(c,r)}$-classes, such that $S_1 \preceq_{(c,r)} S_2$ if an object of $S_1$ is simulated by another object of $S_2$. 

As the first step of investigating structures of $WF$-$\preceq_{(c,r)}$-classes, we consider the problem of 
whether there exist a maximum element and a minimum element of the $WF$-$\preceq_{(c,r)}$ classes. The answers are both yes. 
We generate  
a wait-free object $\mathcal{U}_{Spec}$, and find that the $WF$-$\preceq_{(c,r)}$-class containing $\llbracket \mathcal{U}_{Spec},n \rrbracket$ is just the maximum element of the $WF$-$\preceq_{(c,r)}$-classes, where $\llbracket \mathcal{U}_{Spec},n \rrbracket$ is the operational semantics of $\mathcal{U}_{Spec}$ for $n$ processes. 
We also find that the $WF$-$\preceq_{(c,r)}$-class containing $\llbracket \mathcal{A}_{Spec},n \rrbracket$ is just the minimum element of the  $WF$-$\preceq_{(c,r)}$-classes, where $\mathcal{A}_{Spec}$ is the classical universal construction \cite{DBLP:books/daglib/0020056}. 
Furthermore, we find that each pair $S_1,S_2$ of the  $WF$-$\preceq_{(c,r)}$-classes has a least upper bound, the $WF$-$\preceq_{(c,r)}$-class containing $\llbracket LUB_{(\mathcal{O}_1, \mathcal{O}_2)},n \rrbracket$, with $\mathcal{O}_1 \in S_1$ and $\mathcal{O}_2 \in S_2$. 
To prove this, given objects $\mathcal{O}_1,\mathcal{O}_2$ that are linearizable and wait-free, we generate an object $LUB_{(\mathcal{O}_1, \mathcal{O}_2)}$, and prove that $\llbracket LUB_{(\mathcal{O}_1, \mathcal{O}_2)},n \rrbracket$ is the least upper bound of $\llbracket \mathcal{O}_1,n \rrbracket$ and $\llbracket \mathcal{O}_2,n \rrbracket$ w.r.t. the forward simulation relation. 
Thus, the set of the $WF$-$\preceq_{(c,r)}$-classes forms a bounded join-semilattice w.r.t. the partial order $\preceq_{(c,r)}$. 
Similarly, we prove that such a structure exists for objects linearizable w.r.t. $Spec$ and lock-free (resp., obstruction-free).

If the liveness constraints are of no concern, i.e., it does not matter whether a linearizable object is wait-free, lock-free, or obstruction-free, we then find that the whole set of linearizable objects forms a bounded lattice, instead of a bounded semilattice. The key point is that, we can generate a greatest lower bound object $GLB_{(\mathcal{O}_1,\mathcal{O}_2)}$ for objects $\mathcal{O}_1$ and $\mathcal{O}_2$ w.r.t. the forward simulation relation, but $GLB_{(\mathcal{O}_1,\mathcal{O}_2)}$ violates these liveness constraints, and thus, cannot be contained in any $WF$-$\preceq_{(c,r)}$-class. 
A $\preceq_{(c,r)}$-class is a set of linearizable objects that simulate each other. 
We prove that $\preceq_{(c,r)}$-classes have similar maximum elements and least upper bounds as $WF$-$\preceq_{(c,r)}$-classes. 
We prove that $\llbracket \mathcal{D}_{Spec},n \rrbracket$ is simulated by $\llbracket \mathcal{O},n \rrbracket$ if $\mathcal{O}$ is linearizable w.r.t. $Spec$, where $\mathcal{D}_{Spec}$ is the object where no method call can return. 
Thus, the $\preceq_{(c,r)}$-class containing $\llbracket \mathcal{D}_{Spec},n \rrbracket$ is the minimum element of $\preceq_{(c,r)}$-classes w.r.t. the partial order $\preceq_{(c,r)}$. 
Then, the set of the $\preceq_{(c,r)}$-classes forms a bounded lattice w.r.t. the partial order $\preceq_{(c,r)}$.

Our lattice and semilattice results yield forward-simulation relations between concrete objects. 
We further present two case studies for these simulation relations. 
In the first case study, we demonstrate the power of $\mathcal{U}_{Spec}$ by producing machine-verified proofs of linearizability. 
To be precise, we prove that the Herlihy-Wing queue \cite{DBLP:journals/toplas/HerlihyW90}, which is notorious for being hard to prove correct, is simulated by 
$\mathcal{U}_{Queue}$, where $Queue$ is the sequential specification of queue. 
We formally prove this by explicitly generating the forward simulation relation.

In the second case study, we prove that the Herlihy-Wing queue is simulated by the time-stamped queue \cite{DBLP:conf/popl/DoddsHK15,DBLP:conf/esop/KhyzhaDGP17}, while the time-stamped queue is not simulated by the Herlihy-Wing queue.  
The Herlihy-Wing queue and the time-stamped queue use different approaches to implement the $Queue$ specification, and both are known to be linearizable~\cite{DBLP:conf/popl/DoddsHK15,DBLP:conf/esop/KhyzhaDGP17}. 
To prove that the Herlihy-Wing queue is simulated by the time-stamped queue, we explicitly generate the forward simulation relation.  
The non-simulation result is obtained by generating a concrete execution of a time-stamped queue, such that the last configuration of this execution cannot be simulated by any configuration of the Herlihy-Wing queue. 
These simulation/non-simulation results 
indicate that the implementation manner of the time-stamped queue indeed extends that of the Herlihy-Wing queue. 
Thus, these two objects belong to different $\preceq_{(c,r)}$-classes. 
As far as we know, these are the first simulation/non-simulation results for two concrete, linearizable objects. 
\begin{wrapfigure}{r}{0.42\textwidth}
  \centering
  \includegraphics[width=0.39\textwidth]{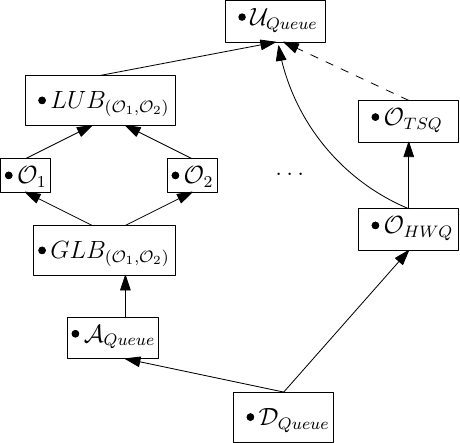} 
  \caption{Lattice hierarchy of queue.} 
  \label{fig:lin lattice}
\end{wrapfigure}
\hspace*{2em}We show the lattice hierarchy of the queue specification $Queue$ in \figurename~\ref{fig:lin lattice}, where each framed node is a $\preceq_{(c,r)}$-class, each directed edge from $\preceq_{(c,r)}$-class $S_1$ to $\preceq_{(c,r)}$-class $S_2$ indicates that $S_1 \preceq_{(c,r)} S_2$, and a dashed edge additionally indicates that this $\preceq_{(c,r)}$ relation has not yet been verified. 
For each $\preceq_{(c,r)}$-class in the figure, we demonstrate a representative object such that the operational semantics of this object is in this $\preceq_{(c,r)}$-class. 
Here $\mathcal{O}_1$ and $\mathcal{O}_2$ are two example queue objects that are linearizable. 
We show that $\mathcal{A}_{Queue}$ (resp., $\mathcal{U}_{Queue}$) simulates $\mathcal{D}_{Queue}$ (resp., $\mathcal{A}_{Queue}$), but not vice versa. 
Moreover, we show that the Herlihy-Wing queue object $\mathcal{O}_{HWQ}$ (resp., the time-stamped queue $\mathcal{O}_{TSQ}$) and $\mathcal{A}_{Queue}$ are in different $\preceq_{(c,r)}$-classes.

The main contributions of this paper are as follows:
\begin{itemize} 
\item[-]  
We discover that the sets of linearizable and wait-free (resp., lock-free, obstruction-free) objects form a bounded join-semilattice, and the sets of linearizable objects form a bounded lattice.

\item[-] We propose an equivalence characterization of linearizability as a forward simulation towards $\llbracket \mathcal{U}_{Spec},n \rrbracket$.

\item[-] We formally verify that the Herlihy-Wing queue is simulated by $\mathcal{U}_{Queue}$. 
This demonstrates that $\mathcal{U}_{Spec}$ is capable of verifying non-trivial objects. 

\item[-] 
We formally prove that the time-stamped queue simulates the Herlihy-Wing queue, while the converse does not hold. 
This offers formal evidence for two linearizable objects located in different $\preceq_{(c,r)}$-classes that are linked by a $\preceq_{(c,r)}$ relation. 
\end{itemize}  

All our case-study proofs have been mechanically checked by Isabelle/HOL~\cite{DBLP:books/sp/NipkowPW02}, and are publicly available at \href{https://github.com/yilyu/lin-lattice}{https://github.com/yilyu/lin-lattice}.\footnote{During proof conceptualization and development, the authors used large language models to brainstorm verification strategies and explore tactical structures. 
Details are provided in the artefact README.}

\subsection{Related Work}
\noindent {\bf Reducing Linearizability to Forward Simulation:} Researchers have investigated the idea of reducing checking linearizability to checking forward simulation towards specific objects. 
Bouajjani \textit{et al.} \cite{DBLP:conf/cav/BouajjaniEEM17} reduced checking linearizability of queue and stack objects into checking forward simulation, and their approach is complete for a specific kind of queue and stack objects. 
Jayanti \textit{et al.} \cite{DBLP:journals/pacmpl/JayantiJYH24} proposed a forward-reasoning-based verification approach by tracking all possible linearizations of ongoing operations. 
Their proof method is also complete for linearizability. 
However, they do not explicitly generate a universal object that simulates a target object if and only if the target object is linearizable. 
Shimon \textit{et al.} \cite{DBLP:journals/corr/abs-2408-11015,DBLP:conf/wdag/Shimon0S24} reduced checking linearizability into checking forward simulation towards a specific labelled transition system (LTS). 
Rather than constructing an LTS, we use an object $\mathcal{U}_{Spec}$.

\noindent {\bf Object Hierarchy:} For a variant of strong linearizability \cite{DBLP:conf/stoc/GolabHW11} of register, Shimon \textit{et al.} \cite{DBLP:journals/corr/abs-2408-11015,DBLP:conf/wdag/Shimon0S24} also provided a sequence of objects, starting from ``atomic register'' and ending at the maximum element w.r.t. a forward simulation relation, such that each object is simulated by its next object (if any) but not vice versa. 
Their idea of investigating the structure of variants of strongly linearizable objects is similar to our idea for linearizable objects, whereas they find a ``linear hierarchy'' and we find a lattice hierarchy in a more general setting.

\noindent {\bf Verification of the Herlihy-Wing queue:} Several previous works have established the correctness of the Herlihy-Wing queue. 
Bouajjani \textit{et al.} \cite{DBLP:conf/cav/BouajjaniEEM17} provide proof of the Herlihy-Wing queue by generating forward simulation towards specific objects. 
Khyzha \textit{et al.} \cite{DBLP:conf/esop/KhyzhaDGP17} proved the linearizability of the Herlihy-Wing queue and time-stamped queue by generating a forward simulation relation towards an object with partially ordered histories. 
Schellhorn \textit{et al.} \cite{DBLP:journals/tocl/SchellhornDW14} proved the linearizability of the Herlihy-Wing queue with backward simulation and interactive theorem prover KIV \cite{Reif1998}. 
Jung \textit{et al.} \cite{DBLP:journals/pacmpl/JungLPRTDJ20} proved 
Herlihy-Wing queue with prophecy variables and prover Coq \cite{rocq-manual}. 
Jayanti \textit{et al.} \cite{DBLP:journals/pacmpl/JayantiJYH24} proved the linearizability of the Herlihy-Wing queue with forward reasoning and prover TLAPS \cite{DBLP:conf/cade/ChaudhuriDLM10}. 
Essentially, they enumerate all possible candidate linearizations and prune these candidates during return actions. 
Our simulation proof of Herlihy-Wing queue w.r.t. $\mathcal{U}_{Queue}$ stores only one linearization at any time, and updates the linearization whenever a new operation takes effect. 
Our approach thus explicitly and clearly states how linearization changes during an execution.

\section{Background}
\label{sec:background}

\subsection{Concurrent Systems and Their Operational Semantics} 
\label{subsec:libraries}

A concurrent object provides several methods through which a client program interacts with the object. 
For simplicity of notations, we assume that a method has just one argument and one return value. 
A most general client is a special client program that exhibits all possible behaviors of an object. 
It simply repeatedly calls an arbitrary method of the object with an arbitrary argument for an arbitrary number of times.

In this paper, we consider concurrent systems of $n$ processes, each running a most general client program to interact with the same concurrent object. 
The operational semantics of a concurrent system can be defined as a $\textit{labeled transition system}$ ($LTS$) $\mathcal{A}=(Q,\Sigma,\rightarrow,q_0)$, where $Q$ is a set of states (a.k.a. configurations, e.g., valuations of control states and registers), $\Sigma$ is an alphabet of transition labels, $\rightarrow\subseteq Q\times\Sigma\times Q$ is a transition relation, and $q_0$ is the initial state.
A finite path of $\mathcal{A}$ is a finite transition sequence $q_0\xrightarrow{\beta_1}q_1\overset{\beta_2}{\longrightarrow}\ldots\overset{\beta_k}{\longrightarrow}q_k$ with $k \! \geq \! 0$, $q_i\in Q$ and $\beta_i\in\Sigma$ for each $1\leq i\leq k$. 
A finite trace of $\mathcal{A}$ is a finite sequence $\beta_1 \cdot \beta_2 \cdot \ldots \cdot \beta_k$ with $k \! \geq \! 0$ and $\beta_i\in\Sigma$ for each $1\leq i\leq k$, if there exists a finite path $q_0\overset{\beta_1}{\longrightarrow}q_1\overset{\beta_2}{\longrightarrow}\ldots\overset{\beta_k}{\longrightarrow}q_k$ of $\mathcal{A}$ with $q_i\in Q$ for each $0\leq i\leq k$. 
The notions of infinite paths and infinite traces of an LTS can be defined similarly and are omitted here. 
The operational semantics of a concurrent system with concurrent object $\mathcal{O}$ and $n$ processes is denoted $\llbracket \mathcal{O}, n\rrbracket$.

\subsection{Linearizability and Liveness}
\label{subsec:linearizability and strong linearizability}

A concurrent object is linearizable if all of its histories are.
A history is a finite sequence of call and return actions, where a return action $\textit{return}(P_{i_1},m_1,a_1)$ matches a call action $\textit{call}(P_{i_2},m_2,a_2)$, i.e., both having the same process identifier (${i_1}={i_2}$) and for the same method ($m_1=m_2$). 
A history is sequential if it starts with a call action, and each call (respectively, return) action is immediately followed by a matching return (respectively, a call) action unless it is the last action. 
A process subhistory $h \vert_i$ with $1\leq i\leq n$ for history $h$ is a history consisting of all and only the actions of process $P_i$ of $h$. 
Two histories $h$ and $h'$ are equivalent, if for each process $P_i$, $h \vert_i=h' \vert_i$.

In a history, let $m(a) \Rightarrow b$ denote an operation of calling method $m$ with argument $a$, which returns value $b$. Thus,
an operation $o$ consists of a pair of a call action, denoted $\textit{inv}(o)$, 
and the follow-up matching return action, denoted $\textit{res}(o)$.  
A sequential specification is a prefix-closed set of finite sequences of operations. 
Given specification $Spec$, let $seqHistory(Spec,n)$ be the set of sequential histories corresponding to $Spec$ for $n$ processes, i.e., $seqHistory(Spec,n)$ =  
$\{ call(P_{i_1},m_1,a_1) \cdot return(P_{i_1},m_1,b_1) \cdot \cdots \cdot call(P_{i_k},m_k,a_k) \cdot return(P_{i_k}$, $m_k,b_k)~\vert~(m_1(a_1) \Rightarrow b_1) \cdot \cdots \cdot (m_k(a_k) \Rightarrow b_k) \in Spec, 
1\leq i_1,\ldots,i_k \leq n\}$. 

A history $h$ induces a happen-before relation $<_h$ over operations, i.e., $o_1 <_h o_2$ if $\textit{res}(o_1)$ occurs before $\textit{inv}(o_2)$ in $h$. 
Then, we recall below the definition of linearizability of \cite{DBLP:journals/toplas/HerlihyW90}. 

\begin{definition}[linearizability 
]
\label{def:linearizability}
A history $h$ is $\textit{linearizable}$ with respect to a sequential specification $Spec$, if there exists an extension $h'$ of history $h$ (by appending zero or more return actions), and a sequential 
history $s\in seqHistory(Spec,n)$, such that

\begin{itemize}
\setlength{\itemsep}{0.5pt}
\item[-] $\textit{complete}(h')$ is equivalent to $s$.
\item[-] For any operations $o_1,o_2$ of $h$, if $o_1 <_h o_2$, then $o_1 <_s o_2$.
\end{itemize}
where $\textit{complete}(h')$ is the maximal subsequence of $h'$ consisting of all the matching call and return actions thereof. Then, $s$ is called a linearization of $h$. 
An object $\mathcal{O}$ is $\textit{linearizable}$ w.r.t. 
$Spec$ for $n$ processes, if for each 
finite trace $e$ of $\llbracket \mathcal{O}, n\rrbracket$, $history(e)$ is linearizable w.r.t. $Spec$, where $history(e)$ is the projection of $e$ into call and return actions.
\end{definition}

When the context is clear, we abuse notation and let an operation sequence, rather than a sequential history, represent a linearization.

We also consider in this paper the liveness properties of concurrent objects, which concern their infinite behaviours. An object $\mathcal{O}$ satisfies a liveness property for $n$ processes if each infinite trace of $\llbracket \mathcal{O}, n\rrbracket$ satisfies the liveness property.  
Wait-freedom \cite{DBLP:journals/toplas/Herlihy91,DBLP:conf/concur/LiangHFS13}, lock-freedom \cite{DBLP:books/daglib/0020056,DBLP:conf/concur/LiangHFS13}, and obstruction-freedom \cite{DBLP:books/daglib/0020056,DBLP:conf/concur/LiangHFS13} are three typical liveness properties. 
It is known that \cite{DBLP:books/daglib/0020056} if an object satisfies wait-freedom (resp., lock-freedom), then it also satisfies lock-freedom (obstruction-freedom).

\subsection{The Classical Universal Construction $\mathcal{A}_{Spec}$}
\label{subsec:the classical universal construction ASpec} 

In this paper, we focus on deterministic, non-blocking sequential specifications. 
A sequential specification $Spec$ can be defined as the set of traces of some LTS, referred to as the LTS of $Spec$, where operations constitute transition labels. 
A sequential specification is deterministic if in its LTS, for each method $m$ and each argument $a$, there is at most one transition with a label 
of method $m$ and argument $a$ from each state $q$. 
A sequential specification is non-blocking if in its LTS, for each method $m$ and each argument $a$, there is at least one transition with a label 
of method $m$ and argument $a$ from each state $q$. 
Widely used specifications, such as those for registers, queues, and stacks, are deterministic. 
Non-blocking guarantees the object $\mathcal{U}_{Spec}$ of the next section to be wait-free. 

Let object $\mathcal{A}_{Spec}$ denote the classical universal construction of Herlihy et al. in \cite{DBLP:books/daglib/0020056}. It maintains a linked list, into which each operation inserts one node. Each node contains information about the operation's method and arguments. Operations across different processes use a consensus object to decide which node to insert into the linked list. A helping mechanism is applied in $\mathcal{A}_{Spec}$ to guarantee wait-freedom. It is known that $\mathcal{A}_{Spec}$ is wait-free and linearizable \cite{DBLP:books/daglib/0020056}.

\section{
Equivalent Characterization of Linearizability with 
$\mathcal{U}_{Spec}$}
\label{sec:universal construction}

In this subsection, given a deterministic and non-blocking sequential specification $Spec$, we generate an object $\mathcal{U}_{Spec}$ and reduce checking linearizability w.r.t. $Spec$ into checking forward simulation w.r.t. $\mathcal{U}_{Spec}$. 

$\mathcal{U}_{Spec}$ uses a variable $his$ to store the history from the beginning to the current time point, uses a variable $lin$ to store a linearization of $his$, uses a set $S_{eff\_ops}$ to store the operations that have already taken effect, and uses an array $S\_var[]$ to store the latest counter for each process. 
Here the combination of process identifier $pid$ and $S\_var[pid]$ is used as a unique operation identity for the current operation of process $P_{pid}$.  

The pseudo code of method $apply()$ of $\mathcal{U}_{Spec}$ is shown in Method \ref{applyOfUniversalConstructionUSpec}. 
At Line 1, we atomically obtain the current process identifier, generate the operation identifier of the current operation, and update $his$ by appending the call action of the current operation. 
At Line 2, we atomically invoke a function $genLin()$ to non-deterministically obtain a linearization of $his$, and insert the current operation into $S_{eff\_ops}$. 
Given a history $h_1$, a set $S_1$ of operations and a operation identifier $id_1$, the function $genLin(h_1, S_1,id_1)$ calculates a linearization $l$ of $h_1$. 
We require that (1) the operation of $id_1$ is in $l$, (2) for each operation $ op \in S_1$, $op$ is in $l$ and the return value of $op$ in $S_1$ is the same as the return value of $op$ in $l$. 
If there is more than one such linearization for $h_1$, then $genLin(h_1, S_1,id_1)$ returns one of them nondeterministically. 
$genLin(his,S_{eff\_ops}$, $oid)$ will finish in a finite number of steps since there are only a finite number of linearizations for $his$ and since $Spec$ is non-blocking. 
At Line 3, we atomically update $his$ by appending the return action of the current operation. 
At Line 4, we atomically update the array $S\_var[]$. 
At Line 5, we return from the current opeaion.

\renewcommand{\algorithmcfname}{Method}
\begin{algorithm}
\KwIn {Invoc $(m,a)$} 
$atomic\{$$his$ := $his \cdot call(pid,oid,m,a)$; $\}$, where $pid$ is the identifier of the current process and $oid := (pid, S\_var[pid])$ \\ 
$atomic\{$ $lin$ := $genLin(his,S_{eff\_ops},oid)$; $S_{eff\_ops} = S_{eff\_ops} \cup \{ (pid,oid,m$, $a,b) \}$ $\}$, where $b$ is the return value of the operation $oid$ in $lin$ \\ 
$atomic\{$$his$ := $his \cdot return(pid,oid,m,b)$;$\}$ \\ 
$atomic\{$S\_var[pid] := S\_var[pid]+1;$\}$ \\ 
\KwRet b; \\ 
\caption{$apply$}
\label{applyOfUniversalConstructionUSpec}
\end{algorithm} 

The following lemma states that $\mathcal{U}_{Spec}$ is linearizable and wait-free.

\begin{lemma}
\label{lemma:U_Spec is linearizable w.r.t Spec for Spec and wait-free}
Given a deterministic and non-blocking sequential specification $Spec$, $\mathcal{U}_{Spec}$ is wait-free for $n$ processes and is linearizable w.r.t. $Spec$ for $n$ processes.
\end{lemma}

Given two LTSs $A_1$ and $A_2$, a forward simulation relation $R$ between configurations of $A_1$ and those 
of $A_2$ can be defined such that (1) the initial configurations of $A_1$ and $A_2$ are in $R$, and (2) if $(c_1,c_2) \in R$, then for any  transition $c_1 {\xrightarrow{\alpha}} c'_1$ in $A_1$, there exists some configuration $c'_2$ reachable from $c_2$ with transitions $l$ such that the visible transitions of $\alpha$ and $l$ are identical, and $(c'_1,c'_2) \in R$. 
Let $A_1 \preceq_{(c,r)} A_2$ denote 
a forward simulation between $A_1$ and $A_2$ where call and return actions are visible, and other actions are considered as internal actions.
Here, `$c$' and `$r$' represent the types of visible actions, i.e., call and return actions. 
$A_1 \preceq_{(c,r)} A_2$ indicates that the behavior of $A_1$ can be simulated by that of $A_2$, and thus, $A_2$ has the ``greater power'' than $A_1$. 

The following theorem states an equivalent characterization of linearizability from the perspective of forward simulations, i.e., $\mathcal{O}$ being linearizable w.r.t. $Spec$ for $n$ processes is equivalent to $\llbracket \mathcal{O}, n$ $\rrbracket \preceq_{(c,r)} \llbracket\mathcal{U}_{Spec}, n  \rrbracket$. 

\begin{theorem}
\label{lemma:Uspec is the maximal element w.r.t prec(c,r), for finite number of processes}
Given a deterministic and non-blocking sequential specification $Spec$ and a concurrent object $\mathcal{O}$, $\mathcal{O}$ is linearizable w.r.t. $Spec$ for $n$ processes, if and only if $\llbracket \mathcal{O}, n \rrbracket \preceq_{(c,r)} \llbracket \mathcal{U}_{Spec}, n \rrbracket$.
\end{theorem}

\section{A Forward Simulation based Lattice for  
Linearizable Objects 
} 
\label{sec:A Weak Simulation-based Hierarchy of Concurrent Objects}

\subsection{A Forward Simulation-based Bounded Join-semilattice for Objects With Liveness Constraints} 
\label{subset: lattice, weak simulation, liveness}

In this subsection, we prove that for the objects that are linearizable w.r.t. $Spec$ and satisfy wait-freedom (resp., lock-freedom, obstruction-freedom), the equivalence classes of their operational semantics form a bounded join-semilattice. 
Let us state the domain, the partial order, the least upper bounds, the maximum and the minimum elements of the bounded join-semilattice.

\noindent\textbf{Domain:} 
Given a deterministic and non-blocking sequential specification $Spec$, a WF-$\preceq_{(c,r)}$-class for $Spec$ is a set $S$ such that (1) each $A\in S$ is an LTS $\llbracket \mathcal{O}, n \rrbracket$ of some concurrent object $\mathcal{O}$, which is linearizable w.r.t. $Spec$ and is wait-free for $n$ processes, (2) for each pair $A_1,A_2\in S$, $A_1 \preceq_{(c,r)} A_2$ and $A_2 \preceq_{(c,r)} A_1$, and (3) for any concurrent object $\mathcal{O}$ that is linearizable w.r.t. $Spec$ and wait-free for $n$ processes, if $\llbracket \mathcal{O},n \rrbracket \preceq_{(c,r)} A$ and $A \preceq_{(c,r)} \llbracket \mathcal{O},n \rrbracket$ for some $A \in S$, then $\llbracket \mathcal{O},n \rrbracket \in S$. 
We can similarly define LF-$\preceq_{(c,r)}$-classes and OF-$\preceq_{(c,r)}$-classes by replacing the requirement of wait-freedom in the definition of WF-$\preceq_{(c,r)}$-classes with that of lock-freedom and obstruction-freedom, respectively. 
Here ``WF'', ``LF'' and ``OF'' indicate wait-freedom, lock-freedom and obstruction-freedom, respectively. 
Recall that if an object satisfies wait-freedom (resp., lock-freedom), then it also satisfies lock-freedom (resp., obstruction-freedom). 

Let $Set_{(WF,Spec)}$, $Set_{(LF,Spec)}$ and $Set_{(OF,Spec)}$ be the set of WF-$\preceq_{(c,r)}$-classes, LF-$\preceq_{(c,r)}$-classes and OF-$\preceq_{(c,r)}$-classes for $Spec$, respectively. 
The domains of the three bounded join-semilattices are $Set_{(WF,Spec)}$, $Set_{(LF,Spec)}$ and $Set_{(OF,Spec)}$, respectively.

\noindent\textbf{Partial Order:} We lift the simulation relation $\preceq_{(c,r)}$ to a relation between two sets $S_1,S_2$ of LTSs as follows. 
We say that $S_1 \preceq_{(c,r)} S_2$ if there exist LTS $A_1 \in S_1$ and LTS $A_2 \in S_2$, such that $A_1 \preceq_{(c,r)} A_2$. 
We prove that $\preceq_{(c,r)}$ is a partial order among WF-$\preceq_{(c,r)}$-classes (resp., LF-$\preceq_{(c,r)}$-classes, OF-$\preceq_{(c,r)}$-classes).

Given two objects $\mathcal{O}_1$ and $\mathcal{O}_2$ that are both linearizable w.r.t. $Spec$ and wait-free for $n$ processes, we say that $(\llbracket \mathcal{O}_1, n  \rrbracket,\llbracket \mathcal{O}_2, n \rrbracket) \in R_{WF}$, if both $\llbracket \mathcal{O}_1,n \rrbracket \preceq_{(c,r)} \llbracket \mathcal{O}_2,n \rrbracket$ and $\llbracket \mathcal{O}_2,n \rrbracket \preceq_{(c,r)} \llbracket \mathcal{O}_1,n \rrbracket$ hold. 
We can similarly define relations $R_{LF}$ and $R_{OF}$ by replacing the requirement of wait-freedom in the definition of $R_{WF}$ with that of lock-freedom and obstruction-freedom, respectively. 
We prove that $R_{WF}$ (resp., $R_{LF}$, $R_{OF}$) is an equivalence relation among operational semantics of concurrent objects that are linearizable w.r.t. $Spec$ and wait-free (resp., lock-free, obstruction-free) for $n$ processes. 
Thus, each WF-$\preceq_{(c,r)}$-class (resp., LF-$\preceq_{(c,r)}$-class, OF-$\preceq_{(c,r)}$-class) is an equivalent class of $R_{WF}$ (resp., $R_{LF}$, $R_{OF}$). 
Therefore, given a concurrent object $\mathcal{O}$ that is linearizable w.r.t. $Spec$ and wait-free (resp., lock-free, obstruction-free) for $n$ processes, $\llbracket \mathcal{O},n \rrbracket$ must belong to a unique WF-$\preceq_{(c,r)}$-class (resp., LF-$\preceq_{(c,r)}$-class, OF-$\preceq_{(c,r)}$-class) for $Spec$.

\noindent {\bf Least Upper Bound:} 
Given two concurrent objects $\mathcal{O}_1$ and $\mathcal{O}_2$ that are linearizable w.r.t. $Spec$, we generate an object $LUB_{(\mathcal{O}_1,\mathcal{O}_2)}$, where ``LUB'' represents least upper bound. 
$LUB_{(\mathcal{O}_1,\mathcal{O}_2)}$ contains an instance $o_1$ of $\mathcal{O}_1$, an instance $o_2$ of $\mathcal{O}_2$ and a new memory location $choice$ with initial value $\bot$. 
$LUB_{(\mathcal{O}_1,\mathcal{O}_2)}$ supports the same set of methods as $\mathcal{O}_1$ and $\mathcal{O}_2$. 
For each method $m$ of $LUB_{(\mathcal{O}_1,\mathcal{O}_2)}$, its pseudo code is shown in Method \ref{LUB(O1,O2)}.  
At Line 1, we read the value of $choice$, and nondeterministically set $choice$ into $0$ or $1$ with the $cas$ command at Line 2 if $choice == \bot$. 
We reread the value of $choice$ in Line 3.
If the value of $choice$ is $0$, then we work as $o_2.m(a)$ at Line 4. 
Otherwise, we work as $o_1.m(a)$ at Line 6. 
Then we return the result of $m(a)$ in Line 7.

\begin{algorithm}
\KwIn {argument $a$} 
\If{$choice == \bot$}
{
    non-deterministically execute $cas(choice,\bot,0)$ or $cas(choice,\bot,1)$; \\ 
} 
\If{$choice == 0$}
{
    $retVal := o_2.m(a)$ ; \\ 
} 
\Else
{
    $retVal := o_1.m(a)$ ; \\ 
} 
\KwRet $retVal$; \\
\caption{method $m$} 
\label{LUB(O1,O2)}
\end{algorithm} 


Given $Prog \in \{ WF,LF,OF \}$ and two $Prog$-$\preceq_{(c,r)}$-classes $S_1,S_2$, let $S_{(S_1,S_2)}^{(lub,Prog)}$ be the unique $Prog$-$\preceq_{(c,r)}$-class that contains $\llbracket LUB_{(\mathcal{O}_1,\mathcal{O}_2)},n \rrbracket$ for some $\llbracket \mathcal{O}_1,n \rrbracket \in S_1$ and $\llbracket \mathcal{O}_2,n \rrbracket \in S_2$. 
The following lemma states that $S_{(S_1,S_2)}^{(lub,Prog)}$ is the least upper bound for each pair $S_1,S_2$ of $Prog$-$\preceq_{(c,r)}$-classes for $Spec$.

\begin{lemma}
\label{lemma:LUB of c,r-classes of linearizable and obstruction-free concurrent objects, for nondeterministic commands and WF, LF, OF} 
Given $Prog \in \{ WF, LF, OF \}$, a deterministic and non-blocking sequential specification $Spec$ and two $Prog$-$\preceq_{(c,r)}$-classes $S_1$ and $S_2$ for $Spec$. 
We have that $S_1 \preceq_{(c,r)} S_{(S_1,S_2)}^{(lub,Prog)}$, $S_2 \preceq_{(c,r)} S_{(S_1,S_2)}^{(lub,Prog)}$. 
Moreover, for each $Prog$-$\preceq_{(c,r)}$-class $S_3$ for $Spec$, if $S_1 \preceq_{(c,r)} S_3$ and $S_2 \preceq_{(c,r)} S_3$, then $S_{(S_1,S_2)}^{(lub,Prog)} \preceq_{(c,r)} S_3$. 
\end{lemma}

\begin{proof} (Sketch) 

First, we prove that $\llbracket LUB_{(\mathcal{O}_1,\mathcal{O}_2)},n \rrbracket$ is the least upper bound of $\llbracket \mathcal{O}_1,n \rrbracket \in S_1$ and $\llbracket \mathcal{O}_2,n \rrbracket \in S_2$ w.r.t. $\preceq_{(c,r)}$. 
Then, we prove that $LUB_{(\mathcal{O}_1,\mathcal{O}_2)}$ is linearizable, and is wait-free (resp., lock-free, obstruction-free) for $n$ processes if both $\mathcal{O}_1$ and $\mathcal{O}_2$ are wait-free (resp., lock-free, obstruction-free) for $n$ processes. 
Finally, the lemma is concluded by ``lifting'' the least upper bound of linearizable objects to the least upper bound of $Prog$-$\preceq_{(c,r)}$-classes. We prove that $\llbracket LUB_{(\mathcal{O}_1,\mathcal{O}_2)},n \rrbracket$ and $\llbracket LUB_{(\mathcal{O}_3,\mathcal{O}_4)},n \rrbracket$ simulate each other for each $\llbracket \mathcal{O}_3,n \rrbracket \in S_1$ and $\llbracket \mathcal{O}_4,n \rrbracket \in S_2$. 
\end{proof}

\noindent {\bf Maximum Element:} 
Let $S_{(\mathcal{U}_{Spec},Prog)}$ be the unique $Prog$-$\preceq_{(c,r)}$-class that contains $\llbracket \mathcal{U}_{Spec},n \rrbracket$ for $Prog \in \{ WF, LF, OF \}$. 
The following lemma states that $S_{(\mathcal{U}_{Spec},Prog)}$ is the maximum element of $Set_{(Prog,Spec)}$.

\begin{lemma}
\label{lemma:maximal element of c,r-classes of linearizable concurrent objects, nondeterministic code and MST} 
Given a deterministic and non-blocking sequential specification $Spec$. 
For each $Prog$-$\preceq_{(c,r)}$-class $S$ for $Spec$ with $Prog \in \{ WF, LF, OF \}$, we have that $S \preceq_{(c,r)} S_{(\mathcal{U}_{Spec},Prog)}$. 
\end{lemma} 

\begin{proof} 
Given a $Prog$-$\preceq_{(c,r)}$-class $S$ for $Spec$ and assume that $\llbracket \mathcal{O},n \rrbracket \in S$. 
Obviously $\mathcal{O}$ is linearizable w.r.t. $Spec$ for $n$ processes. 
By Theorem \ref{lemma:Uspec is the maximal element w.r.t prec(c,r), for finite number of processes} we can see that $\llbracket \mathcal{O},n \rrbracket \preceq_{(c,r)} \llbracket \mathcal{U}_{Spec},n \rrbracket$. 
By Lemma \ref{lemma:U_Spec is linearizable w.r.t Spec for Spec and wait-free} we can see that $\mathcal{U}_{Spec}$ is linearizable and wait-free.  
Thus, we have that $S \preceq_{(c,r)} S_{(\mathcal{U}_{Spec},Prog)}$. 
\end{proof}

\noindent {\bf Minimum Element:} 
Let $S_{(\mathcal{A}_{Spec},Prog)}$ be the unique $Prog$-$\preceq_{(c,r)}$-class that contains $\llbracket \mathcal{A}_{Spec},n \rrbracket$. 
The following lemma states that $S_{(\mathcal{A}_{Spec},Prog)}$ is the minimum element of $Set_{(Prog,Spec)}$.

\begin{lemma}
\label{lemma:minimal element of c,r-classes of linearizable concurrent objects, nondeterministic code and MST} 
Given a deterministic and non-blocking sequential specification $Spec$. 
For each $Prog$-$\preceq_{(c,r)}$-class $S$ for $Spec$ where $Prog \in \{ WF, LF, OF \}$, we have that $S_{(\mathcal{A}_{Spec},Prog)} \preceq_{(c,r)} S$. 
\end{lemma}

The following theorem states that $(Set_{(Prog,Spec)},\preceq_{(c,r)})$ is a bounded join-semila- ttice, which is a direct consequence of Lemma \ref{lemma:LUB of c,r-classes of linearizable and obstruction-free concurrent objects, for nondeterministic commands and WF, LF, OF}, Lemma \ref{lemma:maximal element of c,r-classes of linearizable concurrent objects, nondeterministic code and MST} and Lemma \ref{lemma:minimal element of c,r-classes of linearizable concurrent objects, nondeterministic code and MST}. 

\begin{theorem}
\label{theorem: lattice of linearizable and concurrency-free concurrent objects, MST} 
Given a deterministic and non-blocking sequential specification $Spec$ and $Prog \in \{ WF, LF, OF \}$, $(Set_{(Prog,Spec)}$, $\preceq_{(c,r)})$ is a bounded join-semilattice w.r.t. $\preceq_{(c,r)}$, with maximum element $S_{(\mathcal{U}_{Spec},Prog)}$ and minimum element $S_{(\mathcal{A}_{Spec},Prog)}$.
\end{theorem}

\subsection{A Forward Simulation-based Bounded Lattice for Objects Without Liveness Constraints} 
\label{subsec: A Weak Simulation based Bounded Lattice for Objects Without Liveness Constraint}

In this subsection, we prove that for the objects that are linearizable w.r.t. $Spec$, the equivalent classes of their operational semantics constitute a bounded lattice. 

We define the notion $\preceq_{(c,r)}$-class for $Spec$ from the definition of WF-$\preceq_{(c,r)}$-class for $Spec$ by removing the requirement of wait-freedom. 
Let $Set_{Spec}$ be the set of $\preceq_{(c,r)}$-classes for $Spec$. 
We define the relation $R_{lin}$ from the definition of $R_{WF}$ by removing the requirement of wait-freedom. 
We prove that $\preceq_{(c,r)}$ is a partial order among $\preceq_{(c,r)}$-classes, and $R_{lin}$ is an equivalence relation among operational semantics of concurrent objects that are linearizable w.r.t. $Spec$ for $n$ processes. 
Thus, we let $Set_{Spec}$ be the domain of the lattice, and let $\preceq_{(c,r)}$ be the partial order of the lattice. 
We prove that given two $\preceq_{(c,r)}$-classes $S_1,S_2$, and given $\llbracket \mathcal{O}_1,n \rrbracket \in S_1$ and $\llbracket \mathcal{O}_2,n \rrbracket \in S_2$, we have that $\llbracket LUB_{(\mathcal{O}_1,\mathcal{O}_2)},n \rrbracket$ is a least upper bound of $\llbracket \mathcal{O}_1,n \rrbracket \in S_1$ and $\llbracket \mathcal{O}_2,n \rrbracket \in S_2$ w.r.t. $\preceq_{(c,r)}$.
Let $S_{(S_1,S_2)}^{lub}$ be the unique $\preceq_{(c,r)}$-class that contains $\llbracket LUB_{(\mathcal{O}_1,\mathcal{O}_2)},n \rrbracket$ for some $\llbracket \mathcal{O}_1,n \rrbracket \in S_1$ and $\llbracket \mathcal{O}_2,n \rrbracket \in S_2$. 
Thus, we can see that $S_{(S_1,S_2)}^{lub}$ is the least upper bound for each pair $S_1,S_2$ of $\preceq_{(c,r)}$-classes. 

\begin{algorithm}
\KwIn {argument $a$} 
execute $o_1.call(m,a)$; \\ 
execute $o_2.call(m,a)$; \\ 
\While {$o_1.m(a)$ or $o_2.m(a)$ does not return yet} 
{ 
    non-deterministically execute one command of $o_i$ where $i \in \{0,1\}$ and $o_i$ not return yet; \\ 
} 

\If{$o_1.m(a)$ and $o_2.m(a)$ return the same value}
{
    \KwRet such value; \\ 
} 
\Else 
{
    falls into an infinite loop; \\ 
}
\caption{method $m$} 
\label{GLB(O1,O2)}
\end{algorithm} 

\noindent {\bf Greatest Lower Bound:} 
Given two concurrent objects $\mathcal{O}_1$ and $\mathcal{O}_2$ that are linearizable w.r.t. $Spec$, we generate an object $GLB_{(\mathcal{O}_1,\mathcal{O}_2)}$, where ``GLB'' represents greatest lower bound.  $GLB_{(\mathcal{O}_1,\mathcal{O}_2)}$ contains an instance $o_1$ of $\mathcal{O}_1$ and an instance $o_2$ of $\mathcal{O}_2$. $GLB_{(\mathcal{O}_1,\mathcal{O}_2)}$ supports the same set of methods as $\mathcal{O}_1$ and $\mathcal{O}_2$. 
For each method $m$ of $GLB_{(\mathcal{O}_1,\mathcal{O}_2)}$, its pseudo code is shown in Method \ref{GLB(O1,O2)}. 
At Lines 1 and 2 we execute $o_1.call(m,a)$ and $o_2.call(m,a)$. 
Then, whenever $o_1.call(m,a)$ or $o_2.call(m,a)$ does not return, we nondeterministically schedule them until both return at Line 4. 
If $o_1.m(a)$ and $o_2.m(a)$ return the same value, then we let this $m(a)$ operation of $GLB_{(\mathcal{O}_1,\mathcal{O}_2)}$ return this value at Line 6. 
Otherwise, this $m(a)$ operation falls into an infinite loop at Line 8. 

Let $S_{(S_1,S_2)}^{glb}$ be the unique $\preceq_{(c,r)}$-class that contains $\llbracket GLB_{(\mathcal{O}_1,\mathcal{O}_2)},n \rrbracket$. 
The following lemma states that $S_{(S_1,S_2)}^{glb}$ is the greatest lower bound for each pair $S_1,S_2$ of $\preceq_{(c,r)}$-classes for $Spec$.  

\begin{lemma}
\label{lemma:infimum of c,r-classes of linearizable and obstruction-free concurrent objects, for nondeterministic commands and no livness property constraints} 
Given a deterministic and non-blocking sequential specification $Spec$ and two $\preceq_{(c,r)}$-classes $S_1$ and $S_2$ for $Spec$. 
We have that $S_{(S_1,S_2)}^{glb} \preceq_{(c,r)} S_1$ and $S_{(S_1,S_2)}^{glb}$ $\preceq_{(c,r)} S_2$. 
Moreover, for each $\preceq_{(c,r)}$-class $S_3$ for $Spec$, if $S_3 \preceq_{(c,r)} S_1$ and $S_3 \preceq_{(c,r)} S_2$, then $S_3 \preceq_{(c,r)} S_{(S_1,S_2)}^{glb}$. 
\end{lemma} 

Let $S_{\mathcal{U}_{Spec}}$ be the $\preceq_{(c,r)}$-class that contains $\llbracket \mathcal{U}_{Spec}, n \rrbracket$. 
By Theorem \ref{lemma:Uspec is the maximal element w.r.t prec(c,r), for finite number of processes}, 
we can see that $S_{\mathcal{U}_{Spec}}$ is the maximum element of $Set_{Spec}$ w.r.t. $\preceq_{(c,r)}$. 
Let 
$\mathcal{D}_{Spec}$ be an object that supports methods of $Spec$, but never returns for any of these methods. 
$\mathcal{D}_{Spec}$ is obviously linearizable w.r.t. $Spec$ for $n$ processes, since each history of $\llbracket \mathcal{D}_{Spec},n \rrbracket$ has no return action, and thus, the empty sequence is the linearization for such a history.  
We prove that if an object $\mathcal{O}$ is linearizable w.r.t. $Spec$ for $n$ processes, then $\llbracket \mathcal{D}_{Spec},n \rrbracket \preceq_{(c,r)} \llbracket \mathcal{O},n \rrbracket$. 
Let $S_{\mathcal{D}_{Spec}}$ be the unique $\preceq_{(c,r)}$-class that contains $\llbracket \mathcal{D}_{Spec},n \rrbracket$. 
Thus, $S_{\mathcal{D}_{Spec}}$ is the minimum element of $Set_{Spec}$ w.r.t. $\preceq_{(c,r)}$.  
From the above results, we can see that $(Set_{Spec},\preceq_{(c,r)})$ is a bounded lattice, as shown by the following theorem. 

\begin{theorem}
\label{theorem: lattice of linearizable and concurrency-free concurrent objects, no liveness constraint} 
Given a deterministic and non-blocking sequential specification $Spec$, $(Set_{Spec}$, $\preceq_{(c,r)})$ is a bounded lattice w.r.t. $\preceq_{(c,r)}$, with maximum element $S_{\mathcal{U}_{Spec}}$ and minimum element $S_{\mathcal{D}_{Spec}}$.
\end{theorem}

\noindent {\bf Remarks:} 
\figurename~\ref{fig:lin lattice} illustrates the lattice hierarchy of the queue specification $Queue$. 
The theoretical results of Section \ref{sec:universal construction}, Section \ref{sec:A Weak Simulation-based Hierarchy of Concurrent Objects} and the next Section 
explain most part of \figurename~\ref{fig:lin lattice}. 
We then prove the remaining part, such as $\llbracket \mathcal{O}_{HWQ},n \rrbracket$ and $\llbracket \mathcal{A}_{Queue},n \rrbracket$ being located in different classes. 
We conjecture that $\mathcal{U}_{Queue}$ simulates the time-stamped queue but not vice versa, and this is shown as a dashed edge in the figure.

\section{Case Studies} 
\label{sec:applications} 

In this section, we prove that the Herlihy-Wing queue is simulated by $\mathcal{U}_{Queue}$ (resp., the time-stamped queue). 
but the time-stamped queue is not simulated by the Herlihy-Wing queue.  
These proofs have been checked by the interactive theorem prover Isabelle/HOL \cite{DBLP:books/sp/NipkowPW02}.  

\subsection{Herlihy-Wing Queue is Simulated by 
$\mathcal{U}_{Queue}$} 
\label{subsec:Proving Linearizability of Herlihy-Wing Queue With UQueue} 

\noindent {\bf The Herlihy-Wing queue object:} 
Herlihy-Wing queue 
maintains an array $Q$ with unbounded size and a variable $X$ with initial value $1$. 
$Q$ stores 
queue elements, and $X$ is the index of the next available element. Initially, each element of $Q$ has value $\bot$. 
The pseudo code $enq()$ and $deq()$ methods 
are shown in Method \ref{enqOfHWQInPaper} and Method \ref{deqOfHWQInPaper}, respectively, where $E_1,E_2,E_3,D_1,D_2,D_3,D_4$ are labels of lines of code. $enq(a)$ first obtains the current index and updates it with $i:=getAndInc(X)$, which reads the value of $X$ and increases the value of $X$ by $1$ atomically. 
It then inserts $a$ into $Q$ with $Q[i]:=a$. $deq()$ repeats the loops of Lines 1-3, in each round of which it obtains the current index $l$, and tries to swap a non-$\bot$ element in $Q[1],\ldots,Q[l-1]$. Whenever a non-$\bot$ element is swapped, it is returned in Line 4. 
Here $x:=swap(Q[j],\bot)$ atomically sets $Q[j]$ to $\bot$ and sets $x$ to the 
original value of $Q[j]$.

\begin{figure}[t]
    \centering
    \begin{minipage}[t]{0.36\linewidth}
        \begin{algorithm}[H]
            \SetAlgoLined
            \KwIn{an argument $v \in \mathbb{N}$}
            $E_1$: $i := getAndInc(X)$\; 
            $E_2$: $Q[i] := v$\; 
            $E_3$: \KwRet ack\; 
            \caption{$enq()$ of HWQ}
            \label{enqOfHWQInPaper}
        \end{algorithm}
    \end{minipage}
    \hfill
    \begin{minipage}[t]{0.62\linewidth}
        \begin{algorithm}[H]
            \SetAlgoLined
            \KwIn{no argument}
            $D_1$: $l := X$\; 
            $D_2$: if $l=1$, then go to $D_1$, else $j:=1$\; 
            $D_3$: $x := swap(Q[j],\bot)$. 
            If $x \neq \bot$, then go to $D_4$. 
            Otherwise, if $j = l-1$, then go to $D_1$, else $j := j+1$ and go to $D_3$\; 
            $D_4$: \KwRet $x$\; 
            \caption{$deq()$ of HWQ}
            \label{deqOfHWQInPaper}
        \end{algorithm}
    \end{minipage}
\end{figure}

\forget{ 
\begin{algorithm}[t]
\KwIn {an argument $v \in \mathbb{N}$} 
$E_1$: $i := getAndInc(X)$; \\ 
$E_2$: $Q[i] := v$; \\ 
$E_3$: \KwRet ack; \\ 
\caption{$enq()$ of HWQ}
\label{enqOfHWQInPaper}
\end{algorithm} 


\begin{algorithm}[t]
\KwIn {no argument} 
$D_1$: $l := X$; \\ 
$D_2$: if $l=1$, then go to $D_1$, else $j:=1$; \\ 
$D_3$: $x := swap(Q[j],\bot)$. 
If $x \neq \bot$, then go to $D_4$. 
Otherwise, if $j = l-1$, then go to $D_1$, else $j := j+1$ and go to $D_3$; \\ 
$D_4$: \KwRet $x$; \\ 
\caption{$deq()$ of HWQ}
\label{deqOfHWQInPaper}
\end{algorithm} 
} 

\noindent {\bf The verification of the Herlihy-Wing queue:} 
To prove that the Herlihy-Wing queue is linearizable w.r.t. the queue specification $Queue$, we explicitly generate a forward simulation relation $SimRel\_U$ between the configurations of $\llbracket \mathcal{O}_{HWQ},n \rrbracket$ and the configurations of $\llbracket \mathcal{U}_{Queue},n \rrbracket$.  
Here we consider only data-independent \cite{DBLP:conf/popl/Wolper86} executions, since this simplifies the generation of the history invariants.  

A configuration of $\llbracket \mathcal{O}_{HWQ},n \rrbracket$ (resp., $\llbracket \mathcal{U}_{Queue},n \rrbracket$) stores the control state of each process as well as the valuation of memory locations, and is represented in Isabelle/HOL as a tuple of type $CState$ (resp., $UState$).\forget{Tuples of type $CState$ additionally contain ghost variables $Qback\_arr$ and $V\_var$. 
$Qback\_arr$ stores the values inserted into $Q$, regardless of whether these values are deleted later. 
$V\_var$ is used to ensure the execution is data-independent.} 
The simulation relation $SimRel\_U$ between the Herlihy-Wing queue and $\mathcal{U}_{Queue}$ is a function of type $CState \Rightarrow UState \Rightarrow bool$, and it is defined as a conjunction of invariants. 
These invariants can be further classified into three categories. 
The first category, termed state invariants, constrains the valuation of the control state and data for tuples of type $CState$. 
The second category, called history invariants, constrains the valuation of the $his$ variable 
of tuples of type $UState$. 
The third category, linearization invariants, constrains the valuation of the $lin$ variable of tuples of type $UState$.  
Let type $SysState: CState \times UState$ represent  
a pair consisting of a configuration of $\llbracket \mathcal{O}_{HWQ},n \rrbracket$ and a configuration of $\llbracket \mathcal{U}_{Queue},n \rrbracket$. 
We define a relation $Next: SysState \Rightarrow SysState \Rightarrow bool$. 
$( (cs,us), (cs',us') ) \in Next$ 
specifies how $\mathcal{U}_{Queue}$ performs a corresponding transition from $us$ according to a transition of the Herlihy-Wing queue from $cs$ to $cs'$. 
Two representative examples are shown below, which state that the ``taking-effect-time-point'' of $enq()$ (resp., $deq()$) corresponds to command $E_2$ (resp., command $D_3$ swapping a non-$\bot$ value). 
Given $( (cs,us), (cs',us') ) \in Next$, 
\begin{itemize}
\item[-] 
If $cs$ reaches $cs'$ by executing command $E_2$ of $enq()$, then $us'$ is obtained from $us$ by appending this $enq()$ operation into the tail of $lin$. 

\item[-] 
If $cs$ reaches $cs'$ by executing command $D_3$ of $deq()$ with a non-$\bot$ swap, 
then $us'$ is obtained from $us$ by calculating a new operation sequence $l'$ from $lin$ 
via a function $modify\_lin()$, 
and  
setting $lin$ to be the concatenation of $l'$ and this $deq()$ operation. 
\end{itemize}

We then  prove that $SimRel\_U$ is indeed a simulation between the Herlihy-Wing queue and $\mathcal{U}_{Queue}$, as stated by the following theorem. 
Here, let $\mathcal{O}_{HWQ}$ be the objects of the Herlihy-Wing queue.

\begin{theorem}
\label{lemma:HWQ is simulated by UQueue} 
$\llbracket \mathcal{O}_{HWQ}, n \rrbracket \preceq_{(c,r)} \llbracket \mathcal{U}_{Queue},n \rrbracket$ . 
\end{theorem}

\subsection{Herlihy-Wing Queue is Simulated by Time-stamped Queue} 
\label{subsec:Proving Herlihy-Wing Queue Simulated by Timestamped Queue} 

\noindent {\bf The time-stamped queue object:} In a time-stamped queue, 
each process is associated with a single-producer/multi-consumer pool, and every element of the queue is stored in a pool. 
When process $P_i$ executes $enq(a)$, it adds a node to its pool with value $a$ and the default timestamp $\top$, generates a new timestamp, and attaches it to the node. When process $P_i$ executes $deq()$, it runs a while loop. 
In each round it generates a timestamp $startTs$, scans all the pools and obtains the oldest timestamp of these timestamps (if they are older than $startTS$), 
an tries to remove the node with such oldest timestamp. 
If it successfully removes such node then it exits the loop and returns the value of this node; otherwise, it starts another round. 
Scanning the oldest elements of the pools can be done in a random order to reduce contention. 
Herein, we assume that the 
timestamps are either an integer or $\top$ with $\top$ larger than any integer, as one kind of timestamp used in \cite{DBLP:conf/popl/DoddsHK15}. 
We use the pseudocode from \cite{DBLP:conf/esop/KhyzhaDGP17} and model it as a state transition. 

\noindent  {\bf The simulation relation:} 
Let $\mathcal{O}_{TSQ}$ be the object of the timestamped queue. 
A configuration of $\llbracket \mathcal{O}_{TSQ},n \rrbracket$ stores the control state of each process as well as the valuation of memory locations, and is represented as a tuple of type $TState$. 
The simulation relation $Simulation\_Inv$ between the Herlihy-Wing queue and the time-stamped queue is a function of type $SysState \Rightarrow TState \Rightarrow bool$. 
In $Simulation\_Inv$ 
we use the type $SysState$ instead of $CState$, because the history information of $\llbracket \mathcal{O}_{HWQ},n \rrbracket$ is necessary for generating the simulation relation, and this information can be obtained in the $UState$ component of $SysState$ tuples. 
Since we still need to deal with history invariants of Section \ref{subsec:Proving Linearizability of Herlihy-Wing Queue With UQueue}, we consider only data-independent executions. 
$Simulation\_Inv$ requires that the $SysState$ and the $TState$ components satisfy their respective invariants, and additionally defines the relation $Simulation\_R$ to constrain the valuations of $SysState$ and $TState$ tuples. 
The most critical category of constraints within $Simulation\_Inv$ is used to strictly relate the data in array $Q$ of the Herlihy-Wing queue to the data in pools of the time-stamped queue. 
Some key constraints are as follows: (1) the set of elements already in $Q$ equals the set of elements with non-$\top$ timestamps in the pools, (2) the set of elements that have obtained indexes but are not yet inserted into $Q$ equals the set of elements with $\top$ timestamps in the pools, and (3) the order of indexes for elements in the Herlihy-Wing queue corresponds to the order of timestamps for elements in the time-stamped queue for both aforementioned cases.

For each command label $lab$ of the Herlihy-Wing queue, we generate a lemma $Simulation\_R\_lab$, which proves the simulation relation upon the execution of such a transition by the $CState$ component. 
By combining these lemmas, 
we prove that the Herlihy-Wing queue is simulated by the time-stamped queue, as stated by the following theorem. 

\begin{theorem}
\label{lemma:HWQ is simulated by TSQ} 
$\llbracket \mathcal{O}_{HWQ}, n \rrbracket \preceq_{(c,r)} \llbracket \mathcal{O}_{TSQ}, n \rrbracket$. 
\end{theorem}

\subsection{Time-stamped queue is not simulated by Herlihy-Wing Queue} 
\label{subsec:Proving Timestamped Queue is not Simulated by Herlihy-Wing Queue} 

In this subsection, we prove that the time-stamped queue is not simulated by the Herlihy-Wing queue even when there are only two processes, as stated by the following theorem. 
Thus, the $\preceq_{(c,r)}$-class of the Herlihy-Wing queue and the $\preceq_{(c,r)}$-class of the time-stamped queue are different, and the former is strictly smaller than the latter w.r.t. the $\preceq_{(c,r)}$ order.

\begin{theorem}
\label{lemma:TSQ is not simulated by HWQ} 
$\llbracket \mathcal{O}_{TSQ}, 2 \rrbracket \npreceq_{(c,r)} \llbracket \mathcal{O}_{HWQ}, 2 \rrbracket$. 
\end{theorem} 

\begin{figure}[tbp]
  \centering
  \includegraphics[width=0.9 \textwidth]{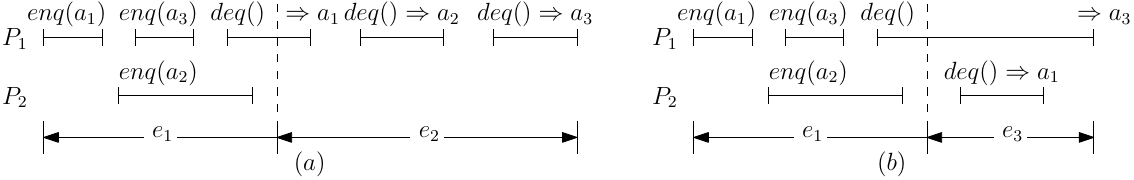} 
\vspace{-5pt}
  \caption{Two executions of $\llbracket \mathcal{O}_{TSQ},2 \rrbracket$ that shares a same prefix $e_1$. }
  \label{fig:TSQ example executions}
\vspace{-5pt}
\end{figure}  
\vspace{-5pt}

We prove it by contradiction and assume that $\llbracket \mathcal{O}_{TSQ}, 2 \rrbracket$ $\preceq_{(c,r)} \llbracket \mathcal{O}_{HWQ}, 2 \rrbracket$. 
To reveal contradiction, we generate a pair of paths of $\llbracket \mathcal{O}_{TSQ},2 \rrbracket$ with the same prefix $e_1$ shown in \figurename~\ref{fig:TSQ example executions}. 
The path $e_1$ is generated as follows: (1) we execute $enq(P_1,a_1)$ until it returns, with $a_1$ associated with time-stamp $ts_1$, (2) we execute $enq(P_2,a_2)$ that inserts a node with value $a_2$ into the pool of process $P_2$, and generates time-stamp $ts_2$ with $ts_1 <_t ts_2$, (3) we execute $enq(P_1,a_3)$ until it returns, with $a_3$ associated with time-stamp $ts_3$ such that $ts_2 <_t ts_3$, (4) we execute $deq(P_1)$ that scans the pool of process $P_2$ and finds no element, and (5) we continue $enq(P_2,a_2)$, associate time-stamp $ts_2$ with $a_2$ and then return. 
Here $<_t$ is the order of the time-stamp. 
In the last configuration $conf$ of $e_1$, the $deq()$ operation $o$ of $e_1$ has not yet scanned the pool of process $P_1$, and finds no element in the pool of process $P_2$. 
In the subsequent execution, if it finds element $a_1$ in the pool of process $P_1$ as shown in the execution $e_2$ of \figurename~\ref{fig:TSQ example executions} (a), then $o$ returns $a_1$. 
In the subsequent execution, if the element $a_1$ is removed by another operation as shown in the execution $e_3$ of \figurename~\ref{fig:TSQ example executions} (b), then $o$ reads $a_3$ from the pool of process $P_1$ and thus returns $a_3$.

We formally modelled both the Herlihy-Wing queue and time-stamped queue using small-step operational semantics, which include visible transitions and $\tau$ transitions. 
We model a predicate $FW\_Sim\_CR$ which is used to determine that there is a forward simulation relation between a configuration of $\llbracket \mathcal{O}_{TSQ},2 \rrbracket$ and a configuration of $\llbracket \mathcal{O}_{HWQ},2 \rrbracket$. 
Based on $e_1$ and the assumption that $\llbracket \mathcal{O}_{TSQ}, 2 \rrbracket$ $\preceq_{(c,r)} \llbracket \mathcal{O}_{HWQ}, 2 \rrbracket$, there should be a configuration $conf'$ such that $(conf,conf')$ satisfies $FW\_Sim\_CR$, where $conf'$ is the last configuration of the corresponding path for $e_1$ in $\llbracket \mathcal{O}_{HWQ},n \rrbracket$. 
We prove that $conf'$ satisfies the property that the set formed by slots 2 and 3 of $Q$ is $\{ a_2, a_3 \}$ (i.e., $\{ Q[2],Q[3] \} = \{ a2, a3 \}$). 
We then prove that to simulate $e_1 \cdot e_2$ with $\llbracket \mathcal{O}_{HWQ},2 \rrbracket$, in $conf'$ we should guarantee that $Q[2]=a_2 \wedge Q[3]=a_3$. 
Finally, we prove that from such $conf'$ there is no subsequent execution with history $history(e_3)$. 
This is because in $conf'$ the element $a_2$ is in $Q[2]$ and thus, the $deq(P_1)$ operation must return $a_2$ instead of $a_3$. 
Thus, $(conf,conf')$ cannot satisfy $FW\_Sim\_CR$ for any $conf'$.

\section{Conclusion and Future Work}
\label{sec:conclusion}

In this paper, we systematically investigate the intrinsic connection between linearizable objects and forward simulation. 
We discover the bounded join-semilattice structure for linearizable and wait-free (resp., lock-free, obstruction-free) objects, and the bounded lattice structure for linearizable objects. 
As far as we know, our lattice/semilattice results are the first to reveal a subtle structure over linearizable objects. 
As part of the lattice/semilattice results, we propose an equivalent characterization of linearizability, which reduces checking whether $\mathcal{O}$ is linearizable w.r.t. $Spec$ for $n$ processes to checking whether $\llbracket \mathcal{O},n \rrbracket$ is simulated by $\llbracket \mathcal{U}_{Spec},n \rrbracket$. 
Our case study with the Herlihy-Wing queue 
demonstrates the potential of $\mathcal{U}_{Spec}$ 
in verification of linearizability. 
 
Our case study of the Herlihy-Wing queue, simulated as a time-stamped queue, demonstrates that objects with different data-manipulation modes can be related through forward simulation. 
It is known that forward simulation preserves hyperproperties \cite{DBLP:journals/jcs/ClarksonS10} for finite traces of client programs in object replacement   \cite{DBLP:conf/wdag/AttiyaE19,DBLP:conf/concur/DongolSW22}. Thus, this example 
promotes future investigation of research on forward simulation between objects. 
Our case study of the time-stamped queue not being simulated by the Herlihy-Wing queue indicates that the former indeed extends the behaviour of the latter. 
To the best of our knowledge, this is the first machine-verified result where two linearizable objects belong to two different elements of the lattice, and such two elements of the lattice are related by the $\preceq_{(c,r)}$ relation.  

A promising line of future work is to further investigate the lattice/semilattice hierarchy. 
For example, it is still unknown if we can generate a greatest lower bound for sets of linearizable and wait-free objects. 
Another direction for future work is to investigate whether such a lattice/semilattice exists for other consistencies.

\forget{ 
\section*{Acknowledgements} During the conceptualisation and development of the formal proofs, the authors utilized the web interfaces of several large language models, including Gemini \cite{gemini2026}, ChatGPT \cite{chatgpt2026}, and DeepSeek \cite{deepseek2026} to brainstorm verification strategies and explore tactical structures. 
All resulting proof scripts were mechanically verified by Isabelle/HOL 
to ensure strict mathematical correctness. 
} 

\bibliographystyle{splncs} 
\bibliography{reference}

\forget{\newpage

\appendix

\input{appendixBackground}

\input{appendixQueryPlus} 

\input{appendixWeakSimulationHierarchy} 

\input{appendixApplication} 

} 

\end{document}